\documentclass[aps,prb,twocolumn,showpacs]{revtex4}
\usepackage{epsf}
 
\begin{document}

       \title{Ultracold atom superfluidity induced 
              by the Feshbach resonance}
      \author{T.\ Domanski}
\affiliation{Institute of Physics, 
             M.\ Curie Sk\l odowska University, 
	     20-031 Lublin, Poland}

\begin{abstract}
We discuss the possible signatures of superfluidity induced 
by the Feshbach resonance in the ultracold gas of fermion 
atoms. Optically or magnetically trapped atoms such as 
$^{6}$Li or $^{40}$K are used in two hyperfine states where 
part of them is converted into the diatomic molecules. These 
fermion and boson entities get coupled in a presence of the 
external magnetic field. Eventually, at critical $T_{c}$, 
they simultaneously undergo transition to the superfluid 
state. Approaching this transition from above there appear 
various signatures manifesting a gradually emerging 
order parameter, but yet the long range coherence is not 
established due to the strong quantum fluctuations. Fermion 
atoms are characterized by the gapped excitation spectrum 
({\em pseudogap}) up to temperature $T_{p}$ (larger than 
$T_{c}$) while boson molecules exhibit collective features 
such as {\em first sound} showing up above a certain 
critical momentum $q_{crit}(T)$. Upon lowering temperature 
to $T_{c}$ this critical value shifts to zero and hence 
there appears the Goldstone mode signaling the symmetry 
broken superfluid state. 
\end{abstract}
\maketitle

\section{Introduction}

During the last fifteen years or so we observe an increased 
experimental and theoretical investigation of the atomic gasses 
which, trapped and cooled to ultralow temperatures ($\sim$ 100 
nK), manifest quantum effects on the macroscopic scale. These 
studies were enabled by a progress of the trapping techniques 
and by unprecedented control of the experimentally adjustable 
interactions between atoms \cite{Ketterle-98}. From the theoretical 
point of view it is important to make a distinction whether 
the number of atom constituents $Z+A$ is odd or even because 
one is confronted either with the fermion or boson objects 
which obey different statistical relations. 

Available quantum states can be occupied by the arbitrary 
number of bosons. Statistical rules enforce that, below critical 
temperature $T_{c}$ bosons start to populate macroscopically the 
lowest lying energy level and this fraction is called the Bose 
Einstein (BE) condensate. Practical realization of such condensates 
has been obtained in the trapped alkali atoms of $^{87}$Rb, 
$^{23}$Na, $^{7}$Li and the polarized hydrogen $^{1}$H. Another 
and the only one naturally existing example of BE condensate is 
$^{4}$He below the $\lambda$ point (i.e.\ under pressure). Strong 
interactions between helium atoms lead there moreover to the 
superfluid behaviour (transport without any observable viscosity). 
Such unique phenomenon can be probably achieved also in the 
trapped alkali atoms where interactions can be experimentally
varied by tuning the external magnetic field.

On the other hand, alkali atoms with odd $Z$ and even $A$  
(such as $^{6}$Li or $^{40}$K) are fermions an must be described 
by the asymmetric wave function as required by the Pauli exclusion 
principle. If the fermion atoms are prepared in two different hyperfine 
states then by switching on the magnetic field their energy levels 
appropriately change and part of the atoms is combined into the 
weakly bound molecules (bosons). These molecules and single atoms 
interact with each other via the resonant type scattering 
\cite{Timmermans-99} from which one can eventually obtain 
the {\em resonant superconductivity} \cite{Kokkelmans-Ohashi}. 
Since atoms are neutral in charge we should rather refer to 
the superfluid instead of superconducting state. In this work 
we shall investigate this boson-fermion mixture with a purpose 
to point out such properties which would enable detection of 
the superfluid transition at $T_{c}$. This issue is extremely
important because the standard experimental methods of the 
condensed matter physics cannot be applied to the trapped atoms.
Moreover, since the superfluid state is smoothly emerging upon 
approaching $T_{c}$ from above we consider also the precursor 
effects which would possibly show up in the experimental 
measurements.

\section{Microscopic description of the Feshbach resonance}

Driving force of the resonant superconductivity/superfluidity 
is a resonant scattering between fermion atoms. Resonances were 
for the first time considered in the atomic physics by Ugo 
Fano and they were later adopted to the nuclear physics by 
Feshbach \cite{Fano-Feshbach}. Starting with a realization 
of such resonances in 1998 \cite{Ketterle-98,first_observation} 
they became a powerful experimental tool for controlling  
the effective scattering potentials ranging between the negative 
to positive values of an arbitrary magnitude.

The resonant Feshbach scattering was recently proposed as 
a mechanism inducing the superconductivity/superfluidity 
of the trapped fermion atoms \cite{Timmermans-99}. On a 
microscopic level the underlying mechanism can be described  
using the following Hamiltonian
\begin{eqnarray}
H &=& \sum_{{\bf k},\sigma} (\varepsilon_{\bf k}-\mu) 
c_{{\bf k}\sigma}^{\dagger} c_{{\bf k}\sigma} 
+ \sum_{\bf q} \left( E_{\bf q} + 2\nu - 2\mu \right) 
b_{\bf q}^{\dagger} b_{\bf q} 
\nonumber \\ & + &
\frac{v}{\sqrt{N}} \sum_{{\bf k},{\bf q}} \left(  
b_{\bf q}^{\dagger} c_{{\bf q}-{\bf k}\downarrow}
c_{{\bf k}\uparrow} + \mbox{h.c.} \right) \nonumber 
\\ & + & \frac{1}{N} \sum_{{\bf k},{\bf p},{\bf q}}
U_{{\bf k},{\bf p}}({\bf q}) c_{{\bf k}\uparrow}^{\dagger} 
c_{{\bf q}-{\bf k}\downarrow}^{\dagger} 
c_{{\bf q}-{\bf p}\downarrow} c_{{\bf p}\uparrow}. 
\label{BF}
\end{eqnarray}
Operators $c_{{\bf k}\sigma}^{(\dagger)}$ refer to the fermion 
atoms in two hyperfine configurations (labeled symbolically by 
$\sigma \! = \! \uparrow$ and $\downarrow$) and $b_{\bf q}^{(\dagger)}$ 
correspond to the diatomic molecules. First terms of the Hamiltonian 
(\ref{BF}) describe kinetic energies of fermions and bosons where 
$\mu$, as usually, denotes the chemical potential. The third term 
describes the coupling between fermion pairs and diatomic 
molecules and the last part denotes a small two-body 
interaction between fermions. The two-body potential 
is often expressed in the atomic physics in terms of 
the scattering length $a$ via the following relation 
$U_{{\bf k},{\bf p}}({\bf q})=\frac{2\pi\hbar^{2}a}{m}n({\bf q})$.

It is the boson-fermion interaction which effectively leads 
to the resonant scattering. In order to prove it in the simplest 
way one can treat $H_{B-F}=\frac{v}{\sqrt{N}} \sum_{{\bf k},{\bf q}} 
\left(  b_{\bf q}^{\dagger} c_{{\bf q}-{\bf k}\downarrow} 
c_{{\bf k}\uparrow} + \mbox{h.c.} \right)$ as a perturbation 
and project out from the Hamiltonian (\ref{BF}) by the canonical 
transformation $e^{S}$. Within the lowest order estimation 
the transformed Hamiltonian becomes \cite{Domanski-pra03} 
\begin{eqnarray}
e^{S}He^{-S} & = & \sum_{{\bf k},\sigma} (\tilde{\varepsilon}
_{\bf k} \! - \! \mu) c_{{\bf k}\sigma}^{\dagger} c_{{\bf k}\sigma} 
+ \sum_{\bf q} \left( \tilde{E}_{\bf q} \! + \! 2\nu \! - \! 2\mu \right) 
b_{\bf q}^{\dagger} b_{\bf q} 
\nonumber \\ & + &
\frac{1}{N} \sum_{{\bf k},{\bf p},{\bf q}}
\tilde{U}_{{\bf k},{\bf p}}({\bf q}) c_{{\bf k}\uparrow}^{\dagger} 
c_{{\bf q}-{\bf k}\downarrow}^{\dagger} 
c_{{\bf q}-{\bf p}\downarrow} c_{{\bf p}\uparrow}, 
\end{eqnarray}
where the two-body potential renormalizes to
\begin{eqnarray}
\tilde{U}_{{\bf k},{\bf p}}({\bf q})=U_{{\bf k},{\bf p}}
({\bf q}) + \frac{v^{2}}{2} \left[ \frac{1}{\varepsilon_
{\bf k} + \varepsilon_{{\bf q}-{\bf k}}-(E_{\bf q}+2\nu)} 
\right. \nonumber \\ +  \left.
\frac{1}{\varepsilon_{\bf p}+\varepsilon_{{\bf q}-{\bf p}}
-(E_{\bf q}+2\nu)} \right] .
\label{resonance}
\end{eqnarray}
In the ${\bf k}={\bf p}$ channel we observe that 
$\tilde{U}_{{\bf k},{\bf p}}({\bf q})$ becomes divergent 
when $\varepsilon_{\bf k}+\varepsilon_{{\bf q}-{\bf k}}
-E_{\bf q}=2\nu$. In particular, considering the atoms 
close to the Fermi energy such divergence occurs for the 
{\em detuning parameter} $\nu = \frac{1}{2} E_{{\bf q}=
{\bf 0}}-\varepsilon_{{\bf k}_F}$. Detuning $\nu$ is 
experimentally adjustable via the external magnetic field. 
Treating $H_{B-F}$ in a better, selfconsistent way the  
divergence of scattering potential is replaced by a finite
resonant-shape jump \cite{Domanski-pra03}.

Resonant Feshbach interactions were already found for the fermion 
atoms of $^{6}$Li (in two hyperfine states $\left| 1/2,1/2 
\right>$, $\left| 1/2,-1/2 \right>$) \cite{Li_atoms} and 
$^{40}$K (in two configurations $\left| 9/2,-9/2\right>$, 
$\left| 9/2,-7/2\right>$) \cite{K_atoms}. Other possible 
realizations are searched for in the heterostructural 
fermion-boson mixtures such as: $^{6}$Li (fermion) with 
$^{7}$Li (boson) \cite{Truscott-01}, $^{6}$Li with 
$^{23}$Na (boson) \cite{Hadzibabic-02}, $^{40}$K 
(fermion) and $^{87}$Rb (boson) \cite{Roati-02}, etc.

\section{Realization of the BCS to BE crossover}

Depending on a value of the detuning parameter $\nu$ there can
occur various kinds of the superconductivity/superfluidity. 
For negative $\nu$ most of the particles are bosons which
at critical $T_{c}$ undergo condensation. The residual 
interactions (of the order $v^{4}$) ultimately induce 
the superfluid state which resembles the BE condensate 
of weakly interacting boson systems. In the opposite limit,  
when $\nu$ is positive and large (say $\nu > v$), boson 
energies are located far above the Fermi level and 
the system consists predominantly of fermions. Virtual 
exchange processes via the boson states generate then  
a kind of the BCS superconductivity of fermions. 

The most interesting situation takes place for the intermediate 
case when $\nu$ is small (positive or negative) because fermions 
and bosons are roughly equally populated and hence their mutual 
interaction is most effective. From the previous studies (see 
for example the review paper \cite{Micnas-90}) it is known that 
transition temperature is optimal under such circumstances. In 
addition to high $T_{c}$ value there arise various symptoms 
(precursor features) of the superfluid order already in the 
normal state. Precursor effects result from the quantum 
fluctuations which, unlike in the usual BCS systems, are 
very strong. Fluctuations manifest up to characteristic 
temperature $T_{p}$ below which fermion pairs are being  
created, yet of only a short life-time.

Since near the Feshbach resonance (for small $\nu$) 
precursor effects play considerable role the usual methods  
for identifying the transition to superconducting/superfluid 
transition in general fail. So far there are three indirect 
indications that superfluidity has been already achieved
among the trapped fermion atoms. These indications are: 
(a) the resonance condensation of fermion pairs 
    \cite{indication_1}, 
(b) a qualitative change of the radial and axial modes 
    of the trapping potential \cite{indication_2}, and 
(c) a double peak structure observed in the radio-frequency 
    spectroscopy \cite{indication_3}.  
First of them is only a necessary condition and is not 
sufficient to confirm the superfluidity. The second point
emphasizes the role of hydrodynamic changes for the cigar
shaped trapping potential \cite{Dalfovo-99}. In a remaining 
part of this work we focus on discussing the third indication 
and eventually also other related experiments whih can help
to infer the superfluid transition.

\section{The RF spectroscopy}

Most of the condensed matter techniques investigating the 
superconducting materials rely upon detecting a gap in 
the single particle spectrum. Temperature at which such 
gap appears is regarded as $T_{c}$. Certainly this 
criterion is not valid here because of pseudogap. 

One of feasible tunneling methods used on the trapped 
atoms is the radio-frequency (RF) spectroscopy 
\cite{RF_spectroscopy}. The main idea is to excite 
selectively one of the fermion species by the appropriately 
adjusted short time ($\sim$ 1s) laser impulses. Let us 
assume that laser is tuned to excite atoms from the state 
$\left| \downarrow \right>$ to another hyperfine state 
which we denote by $\left| e \right>$. Perturbation 
caused by the laser pulse can be described via 
%the following Hamiltonian 
%
\begin{eqnarray}
H_{RF} & = & \sum_{\bf k} \frac{\delta_{RF}}{2} 
\left( c_{{\bf k}e}^{\dagger} c_{{\bf k}e} -
c_{{\bf k}\downarrow}^{\dagger} c_{{\bf k}\downarrow} -
b_{\bf k}^{\dagger} b_{\bf k} \right) \nonumber \\  
& + &\sum_{{\bf k},{\bf p}} \left[ ( M_{{\bf k},{\bf p}}
c_{{\bf k}\downarrow}^{\dagger} c_{{\bf p}e} +
\sum_{\bf q} D_{{\bf k},{\bf p},{\bf q}} 
b_{\bf q}^{\dagger} c_{{\bf k}\uparrow} c_{{\bf p}e}
) + \mbox{h.c} \right], 
\nonumber 
%\label{RF_hamil}
\end{eqnarray}
where the matrix elements $M_{{\bf k},{\bf p}}$,  
$D_{{\bf k},{\bf p},{\bf q}}$ are both proportional to 
the Rabi frequency and the RF detuning parameter is 
$\delta_{RF}=E_{RF}-\varepsilon_{{\bf k}e}-\varepsilon
_{{\bf k}\downarrow}$ with $E_{RF}$ being the photon 
energy. In the experimental setup one is measuring the 
single particle tunneling current of atoms transfered 
to $\left| e \right>$ state what can be expressed by 
the expectation value $I(\delta_{RF})= \langle 
\dot{N}_{e} \rangle$. Because of lack of space we 
do not present the specific expressions for 
$I(\delta_{RF})$ but it can be obtained within 
the linear response theory in a straightforward
manner \cite{Thorma}.

For physical understanding it is enough to explain that 
current $I(\delta_{RF})$ is proportional to the density 
of single particle states $\left| \downarrow \right>$. 
It seems that the recent measurements of Innsbruck group 
on $^{6}$Li indeed provide the first evidence for 
observation of the pairing gap \cite{indication_3}. 
One should keep in mind however, that due to quantum 
fluctuations the gap itself is expected to exist even 
above $T_{c}$ \cite{Domanski_BFmodel} thereof it cannot 
serve as a proof of the superfluid state. Some authors
claimed that asymmetry of the tunneling current would 
be the needed indication of superfluidity. We have 
shown previously \cite{Domanski_PhysicaC} that asymmetry 
is present also in the normal spectrum so this argument 
does not work either.

\section{The Bragg scattering}

The other method applicable for probing the single particle 
gap as well as the correlations between trapped fermions is 
the Bragg spectroscopy. Typical procedure is based on 
weak scattering of the ultracold atoms by a moving potential 
of the form $V_{0} \mbox{cos}({\bf q}{\bf r}-\omega t)$.
Bragg potential can be formed, for instance, by {\em ac} 
Stark shift arising from two interfering laser fields 
\cite{Blakie-02}. Such spectroscopy was pioneered by NIST 
and MIT groups \cite{Bragg_exper} who used it for investigation 
of the BE condensed boson atoms.

Usually the two laser beams used in the experiment are 
polarized and tuned in such a way to allow excitations 
from the state $\left| \downarrow \right>$ to $\left| 
e \right>$. The laser-atom interaction can be described 
by the following Hamiltonian
\begin{eqnarray}
H_{Bragg} & = & \frac{1}{2} V_{0} 
\left( \rho_{\bf q}^{\dagger} \; e^{-i\omega t} +
  \rho_{\bf q} \; e^{i\omega t} \right),
\label{Bragg_hamil}
\end{eqnarray}
where $\rho_{\bf q}=\sum_{\bf k} c_{{\bf k}+{\bf q},
\downarrow}^{\dagger} c_{{\bf k},\downarrow}$ is the 
usual density operator of $\downarrow$ atoms. Bragg 
spectroscopy is sensitive to the density-density 
correlation function analyzed by the time-of-flight 
techniques. Thus measured {\em dynamic structure 
factor} is given by $S_{\rho}({\bf q},\omega)=\frac{1}
{\cal{Z}} \sum_{i,j} e^{-E_{i}/k_{B}T}\left| \left< 
i \right| \rho_{\bf q} \left| j \right> \right| ^{2} 
\delta \left( \omega - E_{j}+E_{i} \right)$, where 
$\cal{Z}$ is the partition function and $E_{i}$ denote 
eigenenergies of the unperturbed Hamiltonian (\ref{BF}).
For the ground state of BCS superconductor  
\cite{Schrieffer_book} one finds \cite{Deb-04}
\begin{eqnarray}
S_{\rho}^{BCS}({\bf q},\omega)=\sum_{\bf k} 
\left| u_{{\bf k}+{\bf q}} \; v_{\bf k} \right| ^{2} 
\delta \left( \omega - E_{{\bf k}+{\bf q}}-E_{\bf k} \right),
\label{S_factor}
\end{eqnarray}
where $|u_{\bf k}|^{2}, |v_{\bf k}|^{2} =\frac{1}{2}[1 \pm 
\xi_{\bf k}/E_{\bf k}]$ with $\xi_{\bf k}=\varepsilon_{\bf k}-\mu$ 
and $E_{\bf k}=\sqrt{\xi_{\bf k}^{2}+\Delta^{2}}$. Bragg pulses 
are thus absorbed only at frequencies $\omega \geq 2 \Delta$  
so this method can measure value of the gap. In practice  
expression (\ref{S_factor}) should be modified taking into
account finite temperature $T \neq 0$ and the effect of 
spatial variation of the trapping potential $V_{trap}=
\frac{m}{2} \left[ \omega_{\perp} (x^{2}+y^{2}) + 
\omega_{||} z^{2} \right]$. The sharp absorption edge 
is then replaced by a kink occurring in the function 
$S_{\rho}({\bf q},\omega)$ at energy $\omega =2 \mbox{max} 
\left\{ \Delta({\bf r}) \right\}$  (see figure 1 in the 
Ref.\ \cite{Deb-04}). 

Besides measuring a value of the single particle gap 
Bragg spectroscopy can also detect some collective 
features. Density-density correlation function $\langle 
\rho_{\bf q}(t) \rho_{\bf q}(0)\rangle$ was shown 
\cite{Tomek-96,Ohashi-03} to be convoluted with the phase 
and amplitude fluctuations of the order parameter. Collective 
phase oscillations ({\em phasons}) are characterized below 
$T_{c}$ by the famous Goldstone mode \cite{Anderson-63}. 
This mode shows up in the dynamic structure factor $S_{\rho}
({\bf q},\omega)$ as a narrow peak appearing in the long
wave-length limit ${\bf q} \rightarrow {\bf 0}$. Such 
property could be used in the future studies as a tool 
for identifying the superfluid state. 

\section{Pair excitations}

Probably the most unambiguous way for investigating 
the transition to superfluid state might be achieved by 
analysis of the pair excitation spectrum. Spectral function 
of the fermion pair operator $\pi_{\bf q}^{\dagger}=\frac{1}{N} 
\sum_{\bf k} c_{{\bf q}-{\bf k}\uparrow}^{\dagger}c_{{\bf k} 
\downarrow}^{\dagger}$ is defined as $S_{\pi}({\bf q},\omega)
=\frac{1}{\cal{Z}} \sum_{i,j} e^{-E_{i}/k_{B}T}\left| \left< 
i \right| \pi_{\bf q} \left| j \right> \right| ^{2} \delta 
\left( \omega - E_{j}+E_{i} \right)$. Close to the Feshbach 
resonance (i.e.\ for $\nu \sim 0$) this spectral function 
was shown \cite{Domanski_tobepubl} to become 
\begin{eqnarray}
S_{\pi}({\bf q},\omega) & = & W^{coh}_{\bf q} \;\; \delta \left[ 
\omega \! - \! (\tilde{E}_{\bf q}\! + \!2\nu \!-\!2\mu) \right]
\label{pair_spectral} \\ & + &
\frac{1}{N} \sum_{\bf k}  W^{inc}_{{\bf q},{\bf k}} \;\;
\delta \left[ \omega \! - \! (\tilde{\varepsilon}_{\bf k} \!
-\!\mu) \! - \! (\tilde{\varepsilon}_{{\bf q}-{\bf k}}\!-\!\mu) 
\right]. \nonumber
\end{eqnarray}
$W^{coh}_{\bf q}$ and $\sum_{\bf k} W^{inc}_{{\bf q},
{\bf k}}$ are the weights of coherent and incoherent parts 
in the fermion pair spectrum which along with renormalized 
energies $\tilde{E}_{\bf k}$, $\tilde{\varepsilon}_{\bf k}$ 
were obtained by the continuous diagonalization 
procedure \cite{Domanski_BFmodel}. 

%    ----------   Figure 1   ----------    %
\begin{figure}
\centerline{\epsfxsize=8cm \epsffile{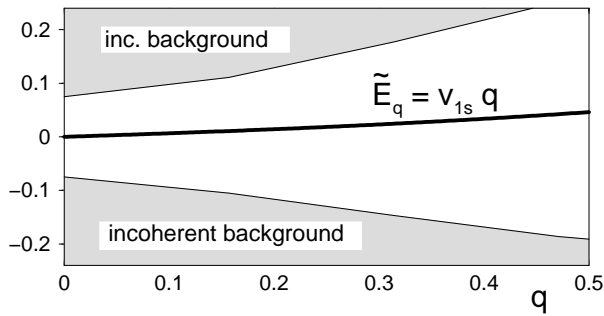}}
\caption{Fermion pair excitation spectrum
in the ground state of the superfluid phase.}
\end{figure}  

%    ----------   Figure 2   ----------    %
\begin{figure}
\centerline{\epsfxsize=8cm \epsffile{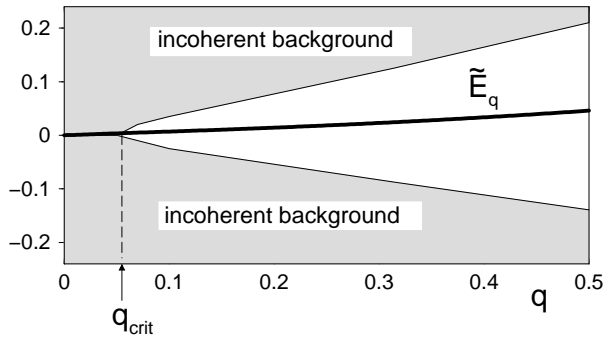}}
\caption{Spectrum of the fermion pair excitations
in the pseudogap regime for $T=0.004$.}
\end{figure}  

In the superfluid state the single particle fermion 
energies $\tilde{\varepsilon}_{\bf k}$ become gapped 
around the Fermi level and hence the incoherent 
part of the pair spectrum (\ref{pair_spectral}) forms 
only outside the energy window $|\omega| \geq 2 \Delta_{sc}$. 
The coherent part on the other hand is characterized by 
a gapless linear ({\em first sound}) mode $lim_{{\bf q}
\rightarrow {\bf 0}}\tilde{E}_{\bf q}+2 \nu-2\mu \propto 
| {\bf q}|$ which appears at small energies as a narrow peak 
in $S_{\pi}({\bf q},\omega)$. Since incoherent background 
is expelled to $\omega>2\Delta_{sc}$ this coherent branch 
becomes well detectable. Existence of such Goldstone mode 
signifies the broken symmetry caused by the order parameter 
$\langle c_{-{\bf k}\downarrow}c_{{\bf k}\uparrow} \rangle 
\neq 0$ but in the charged superconducting systems could 
not be observed due to the long range Coulomb interactions 
lifting it to plasma frequencies \cite{Anderson-63}. 
The neutral fermion atom gasses are very good candidates 
for this mode to be observed. Its observation would 
prove achievement of the atom superfluidity. 

In the normal state the Goldstone mode fades away. For  
low momenta the pair dispersion $\tilde{E}_{\bf q}$
becomes parabolic (massive) and moreover overlaps with
the incoherent background. It is thereof hardly visible
at all. However, at sufficiently large momenta exceeding 
critical value $q_{crit}(T)$ the coherent part splits 
off from the incoherent background and again there appears 
a remnant of the linear branch \cite{Domanski_prlprb}. 
Sound velocity $v_{1s}$ of the colletive branch seems to 
be rather independent of temperature what is typical for the 
strongly interacting boson systems \cite{Szepfalusy-74}.  

%    ----------   Figure 3   ----------    %
\begin{figure}
\centerline{\epsfxsize=3cm \epsffile{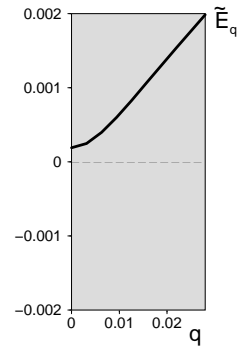}}
\caption{Low energy part of the pair excitation spectrum
in the pseudogap regime.}
\end{figure}

\section{Summary}
We studied some signatures of superfluidity possible 
to induce in the trapped fermion atoms by the Feshbach 
resonance. We claim that, unlike in the usual BCS systems, 
one cannot rely upon appearance of the single particle 
gap as a criterion for $T_{c}$ because such gap exists 
there even in the normal state. Experimental methods 
should focus, in our opinion, on analysis of other 
many-body effects. For instance the pair excitation 
spectrum is expected to undergo qualitative changes
near transition temperature. Appearance of pseudogap 
in the normal state is usually accompanied by emergence 
of the fermion pairs \cite{Domanski_BFmodel}
whose life-time gradually increases for temperature 
approaching $T_{c}$. The pair excitations reveal 
a remnant of the collective first sound
at sufficiently large momenta $q>q_{crit}(T)$ 
\cite{Domanski_prlprb}. This feature can be detected 
in practice by the Bragg spectroscopy. For $T \leq T_{c}$ 
fermion pairs acquire  
the infinite life-time and simultaneously their coherence 
spreads over the macroscopic (long-range) distance. In 
consequence, 
the collective mode extends down to the zero $q_{crit}
(T \! < \! T_{c}) \! = \! 0$. This property is one 
of possible ways for determination of the transition 
temperature.

\vspace{0.1cm}
Author kindly acknowledges collaboration and instructive 
discussions with J.\ Ranninger on the problems discussed
in this work and on the other related issues. This work is supported 
by the Polish Committee of Scientific Research under 
the grant No.\ 2P03B06225.

\end{document}